\documentclass[conference]{IEEEtran}

\usepackage{cite}
\usepackage{amsmath}
\usepackage{amssymb}
\usepackage{amsfonts}
\usepackage{graphicx}
\usepackage[dvipsnames]{xcolor}
\usepackage[a4paper, total={184mm,239mm}]{geometry}
\usepackage[english]{babel}
\usepackage[babel=true,expansion=alltext,protrusion=alltext-nott,final]{microtype}
\usepackage{import}
\usepackage[free-standing-units,per-mode=repeated-symbol,binary-units=true,detect-weight=true,detect-family=true,retain-explicit-plus=true]{siunitx}
\usepackage{glossaries}
\usepackage[capitalise,nameinlink,noabbrev]{cleveref}
\usepackage{pgfplots}
\usepackage{pgfplotstable}
\usepackage{tikz}
\usepackage{booktabs}
\usepackage{textcomp}
\usepackage{enumitem}
\usepackage[hyphens]{url}
\usepackage{nth}
\usepackage[version=4]{mhchem}
\usepackage{newtxmath}
\usepackage{multirow}
\usepackage{stfloats}

\usetikzlibrary{scopes, calc, shapes, arrows, patterns, positioning}
\tikzset{>=latex}

\pgfplotsset{compat=1.13}
\pgfplotsset{width=\linewidth, height=7cm}
\pgfplotsset{every x tick label/.append style={font=\small}}
\pgfplotsset{every y tick label/.append style={font=\small}}

\makeatletter
\let\MYcaption\@makecaption
\makeatother
\usepackage[font=footnotesize]{subcaption}
\makeatletter
\let\@makecaption\MYcaption
\makeatother

\makeatletter
\g@addto@macro{\UrlBreaks}{\UrlOrds}
\makeatother

\renewcommand{\baselinestretch}{0.99}

\DeclareSIUnit\bits{bits}
\DeclareSIUnit\bps{bps}
\DeclareSIUnit\Bps{Bps}
\DeclareSIUnit\core{core}
\DeclareSIUnit\tile{tile}
\DeclareSIUnit\request{request}
\DeclareSIUnit\cycle{cycle}
\DeclareSIUnit\erlang{E}
\DeclareSIUnit\flop{FLOP}
\DeclareSIUnit\flops{FLOPS}
\DeclareSIUnit\gate{GE}
\DeclareSIUnit\op{OP}
\DeclareSIUnit\ops{OPS}
\DeclareSIUnit\ipc{IPC}

\definecolor{color1}{HTML}{256DFF}
\definecolor{color2}{HTML}{45CCB0}
\definecolor{color3}{HTML}{9775CA}
\definecolor{color4}{HTML}{C83737}
\definecolor{color5}{HTML}{77A4FF}
\colorlet{colorAlert}{Red}

\newacronym[longplural={Scratchpad Memories}]{SPM}{SPM}{Scratchpad Memory}
\newacronym{ACE}{ACE}{AXI Coherent Extensions}
\newacronym{AMBA}{AMBA}{Advanced Microcontroller Bus Architecture}
\newacronym{APB}{APB}{Advanced Peripheral Bus}
\newacronym{API}{API}{Application Programming Interface}
\newacronym{ASIC}{ASIC}{Application-Specific Integrated Circuit}
\newacronym{AVX}{AVX}{Advanced Vector Extension}
\newacronym{AXI}{AXI}{Advanced eXtensible Interface}
\newacronym{BLAS}{BLAS}{Basic Linear Algebra Subprograms}
\newacronym{CHI}{CHI}{Coherent Hub Interface}
\newacronym{CMOS}{CMOS}{Complementary Metal-Oxide-Semiconductor}
\newacronym{CNN}{CNN}{Convolutional Neural Network}
\newacronym{CPU}{CPU}{Central Processing Unit}
\newacronym{CSR}{CSR}{Control and State Register}
\newacronym{CTS}{CTS}{Clock Tree Synthesis}
\newacronym{DLP}{DLP}{Data Level Parallelism}
\newacronym{DMA}{DMA}{Direct Memory Access}
\newacronym{DRAM}{DRAM}{Dynamic Random-Access Memory}
\newacronym{DSP}{DSP}{Digital Signal Processing}
\newacronym{DUT}{DUT}{Device Under Test}
\newacronym{ECL}{ECL}{Emitter-Coupled Logic}
\newacronym{FBB}{FBB}{Forward Body-Biasing}
\newacronym{FDSOI}{FD-SOI}{Fully Depleted Silicon on Insulator}
\newacronym{FMA}{FMA}{Fused Multiply-Add}
\newacronym{FPGA}{FPGA}{Field-Programmable Gate Array}
\newacronym{FPU}{FPU}{Floating Point Unit}
\newacronym{GPGPU}{GPGPU}{General-Purpose \acrlong{GPU}}
\newacronym{GPU}{GPU}{Graphics Processing Unit}
\newacronym{HDL}{HDL}{Hardware Description Language}
\newacronym{HERO}{HERO}{Heterogeneous Embedded Research Platform}
\newacronym{HPC}{HPC}{High-Performance Computing}
\newacronym{ILP}{ILP}{Instruction Level Parallelism}
\newacronym{IOT}{IoT}{Internet-of-Things}
\newacronym{IPC}{IPC}{Instructions Per Cycle}
\newacronym{IPU}{IPU}{Image Processing Unit}
\newacronym{ISA}{ISA}{Instruction Set Architecture}
\newacronym{LSU}{LSU}{Load/Store Unit}
\newacronym{LVT}{LVT}{low voltage threshold}
\newacronym{MIMD}{MIMD}{multiple instruction, multiple data}
\newacronym{MMU}{MMU}{Memory Management Unit}
\newacronym{MUL}{MUL}{multiplier}
\newacronym{MVL}{MVL}{maximum vector length}
\newacronym{NUMA}{NUMA}{non-uniform memory access}
\newacronym{NOC}{NoC}{Network-on-Chip}
\newacronym{PCIe}{PCIe}{Peripheral Component Interconnect Express}
\newacronym{PC}{PC}{Program Counter}
\newacronym{PE}{PE}{processing element}
\newacronym{PL}{PL}{Programmable Logic}
\newacronym{PMCA}{PMCA}{Programmable Manycore Accelerator}
\newacronym{PSL}{PSL}{Power Service Layer}
\newacronym{PTE}{PTE}{page-table entry}
\newacronym{PTW}{PTW}{page-table walker}
\newacronym{PULP}{PULP}{Parallel Ultra Low Power}
\newacronym{RAW}{RAW}{read-after-write}
\newacronym{RBB}{RBB}{Reverse Body-Biasing}
\newacronym{ROB}{ROB}{Reorder Buffer}
\newacronym{RTL}{RTL}{Register Transfer Level}
\newacronym{RVT}{RVT}{Regular Voltage Threshold}
\newacronym{RoCC}{RoCC}{Rocket Custom Coprocessor Interface}
\newacronym{SCM}{SCM}{Storage Class Memory}
\newacronym{SIMD}{SIMD}{single instruction, multiple data}
\newacronym{SIMT}{SIMT}{single instruction, multiple thread}
\newacronym{SLDU}{SLDU}{Slide Unit}
\newacronym{SLVT}{SLVT}{super-low voltage threshold}
\newacronym{SM}{SM}{Streaming Multiprocessor}
\newacronym{SRAM}{SRAM}{Static Random-Access Memory}
\newacronym{SSE}{SSE}{Streaming SIMD Extension}
\newacronym{SVE}{SVE}{Scalable Vector Extension}
\newacronym{TLP}{TLP}{Thread Level Parallelism}
\newacronym{TxnID}{TxnID}{Transaction ID}
\newacronym{VAC}{VAC}{Vector Access}
\newacronym{VC}{VC}{virtual channel}
\newacronym{VCONV}{VCONV}{Vector Conversion}
\newacronym{VEX}{VEX}{Vector Execute}
\newacronym{VFU}{VFU}{vector functional unit}
\newacronym{VID}{VID}{Vector Instruction Decode}
\newacronym{VIS}{VISSUE}{Vector Instruction Issue}
\newacronym{VLIW}{VLIW}{Very Long Instruction Word}
\newacronym{VLOOP}{VLOOP}{Vector Loop}
\newacronym{VLR}{VLR}{vector length register}
\newacronym{VLSU}{VLSU}{Vector Load/Store Unit}
\newacronym{VNB}{VNB}{Von Neumann Bottleneck}
\newacronym{VRF}{VRF}{Vector Register File}
\newacronym{VT}{VT}{vector thread}
\newacronym{WAR}{WAR}{write-after-read}
\newacronym{WAW}{WAW}{write-after-write}
\newacronym{DCT}{DCT}{discrete cosine transform}
\newacronym{TSV}{TSV}{through-silicon via}
\newacronym{3DIC}{3D-IC}{three-dimensional integrated circuit}
\newacronym{PPA}{PPA}{power, performance, and area}
\newacronym{F2F}{F2F}{face-to-face}
\newacronym{W2W}{W2W}{wafer-to-wafer}
\newacronym{IC}{IC}{integrated circuit}
\newacronym{C4}{C4}{controlled collapse chip connection}
\newacronym{FEOL}{FEOL}{front end of the line}
\newacronym{BEOL}{BEOL}{back end of the line}
\newacronym{PDP}{PDP}{power-delay product}
\newacronym{EDP}{EDP}{energy-delay product}
\newacronym{DRV}{DRV}{design rule violation}
\newacronym{DDR}{DDR}{double data rate}
\newacronym{SDRAM}{SDRAM}{synchronous dynamic random-access memory}

\newcommand\eg{e.g.,\ }
\newcommand\ie{i.e.,\ }
\newcommand\etal{et\penalty50\ al.\ }
\newcommand\mempool[2]{$\text{MemPool-#2}_{\text{\SI{#1}{\mebi\byte}}}$}
\newif\ifreviewmode

\begin{document}

\title{MemPool-3D: Boosting Performance and Efficiency of Shared-L1 Memory Many-Core Clusters with\\3D Integration}
\ifreviewmode
\author{\emph{Hidden for double-blind review purposes.}}
\else
\author{\IEEEauthorblockN{Matheus Cavalcante\IEEEauthorrefmark{1}\kern-.12em, Anthony Agnesina\IEEEauthorrefmark{2}\kern-.1em, Samuel Riedel\IEEEauthorrefmark{1}\kern-.12em, Moritz Brunion\IEEEauthorrefmark{3}\kern-.1em,\\
                          Alberto García-Ortiz\IEEEauthorrefmark{3}\kern-.1em, Dragomir Milojevic\IEEEauthorrefmark{4}\kern-.1em, Francky Catthoor\IEEEauthorrefmark{4}\kern-.1em, Sung Kyu Lim\IEEEauthorrefmark{2}\kern-.1em, and Luca Benini\IEEEauthorrefmark{1}\kern-.08em\IEEEauthorrefmark{5}}
  \IEEEauthorblockA{\IEEEauthorrefmark{1}ETH Zürich, Zürich, Switzerland}
  \IEEEauthorblockA{\IEEEauthorrefmark{2}Georgia Institute of Technology, Atlanta, Georgia, USA}
  \IEEEauthorblockA{\IEEEauthorrefmark{3}Universität Bremen, Bremen, Germany}
  \IEEEauthorblockA{\IEEEauthorrefmark{4}IMEC, Leuven, Belgium}
  \IEEEauthorblockA{\IEEEauthorrefmark{5}Università di Bologna, Bologna, Italy}
  \IEEEauthorblockA{Email: matheus.cavalcante \emph{at} iis.ee.ethz.ch}}
\fi
\maketitle

\begin{abstract}
  Three-dimensional integrated circuits promise power, performance,
  and footprint gains compared to their 2D counterparts, thanks to
  drastic reductions in the interconnects' length through their
  smaller form factor. We can leverage the potential of 3D integration
  by enhancing MemPool, an open-source many-core design with 256 cores
  and a shared pool of L1 scratchpad memory connected with a
  low-latency interconnect. MemPool's baseline 2D design is severely
  limited by routing congestion and wire propagation delay, making the
  design ideal for 3D integration. In architectural terms, we increase
  MemPool's scratchpad memory capacity beyond the sweet spot for 2D
  designs, improving performance in a common digital signal processing
  kernel.  We propose a 3D MemPool design that leverages a smart
  partitioning of the memory resources across two layers to balance
  the size and utilization of the stacked dies. In this paper, we
  explore the architectural and the technology parameter spaces by
  analyzing the power, performance, area, and energy efficiency of
  MemPool instances in 2D and 3D with \SIlist{1;2;4;8}{\mebi\byte} of
  scratchpad memory in a commercial \SI{28}{\nano\meter} technology
  node. We observe a performance gain of \SI{9.1}{\percent} when
  running a matrix multiplication on the MemPool-3D design with
  \SI{4}{\mebi\byte} of scratchpad memory compared to the MemPool 2D
  counterpart. In terms of energy efficiency, we can implement the
  MemPool-3D instance with \SI{4}{\mebi\byte} of L1 memory on an
  energy budget \SI{15}{\percent} smaller than its 2D counterpart, and
  even \SI{3.7}{\percent} smaller than the MemPool-2D instance with
  one-fourth of the L1 scratchpad memory capacity.
\end{abstract}


\IEEEpeerreviewmaketitle

\glsresetall


\section{Introduction}
\label{sec:introduction}

Vertical integration promises to address the scaling problems of the
traditional 2D integration foreseen by Moore's
Law~\cite{Panth2013}. \Glspl{3DIC} promise better \gls{PPA} than 2D
counterparts, thanks to a drastic reduction of the interconnect
lengths, particularly of long global interconnects, while enabling a
smaller form factor by adding the third dimension~\cite{Pavlidis2009,
  Dong2010}.

Advances in flip-chip interconnection technology allow for the
miniaturization of the inter-die connections. While \gls{C4} solder bumps
have a pitch of around \SI{100}{\micro\meter}~\cite{Tsai2017},
\gls{F2F} wafer-to-wafer hybrid bonding enables interconnect pitches in the
micrometer range while maintaining reasonable yield
rates~\cite{Beyne2017}. Such a fine pitch can be leveraged to
implement \gls{3DIC} designs with a very high interconnect density.

For the implementation of \gls{F2F}-bonded \glspl{3DIC}, the Macro-3D
flow~\cite{Bamberg2020} provides state-of-the-art \gls{PPA}
optimization capabilities for memory-on-logic partitioning schemes.
Since the flow is aware of all metal layers in the die stack, the
\gls{BEOL} routing resources of both dies can be
shared~\cite{Bamberg2020}. It is, therefore, possible to use one
chip's \gls{BEOL} to avoid congestion bottlenecks in the other
chip. This resource sharing allows for a more efficient routing
utilization, which is extremely useful for the implementation of
highly congested designs.

Many-core systems achieve better \gls{PPA} through vertical
integration due to the shorter interconnect lengths. Although some
many-core systems explore \gls{3DIC}
implementations~\cite{Dong2010,Healy2010,Sai2019}, they use a 2D-mesh
network to connect their processing elements, failing to exploit the
interconnection capabilities of a denser pitch. In this paper, we use
MemPool~\cite{MemPool2021} as our target design, an
open-source~\cite{MemPoolGitHub2021} many-core system with \num{256}
cores and a configurable amount of shared L1 \gls{SPM}
connected with a low-latency interconnect. Routing congestion severely
limits MemPool's implementation, with its operating frequency bounded
by the wire propagation delay. This makes MemPool an ideal candidate
for 3D design. With the holistic view over all metal layers in the
Macro-3D flow, the \gls{BEOL} resources in both dies are combined,
alleviating MemPool's congestion and achieving a higher
operating frequency thanks to reductions in the wire length.  In this
paper, we explore the architectural and the technology parameter
spaces by analyzing the \gls{PPA} impact of MemPool's L1 \gls{SPM}
capacity scaling, from \SIrange{1}{8}{\mebi\byte}, on 2D and 3D
implementations. The contributions of this paper are:
\begin{itemize}
\item A flexible partitioning scheme of MemPool into logic and memory
  dies, capable of achieving high utilization of the memory die for
  large memory capacities (\Cref{sec:tile-implementation});
\item The complete 2D and Macro-3D implementations of MemPool in a
  commercial \SI{28}{\nano\meter} technology node, for all considered
  \gls{SPM} capacities, and an analysis of the instances in terms of
  power, performance, area, footprint, and energy efficiency
  (\Cref{sec:tile-implementation,sec:group-impl-results});
\item An exploration of MemPool's \gls{SPM} capacity and its impact on
  the runtime of a common matrix multiplication kernel, including an
  analysis of the off-chip memory bandwidth's influence
  (\Cref{sec:mempools-performance}).
\end{itemize}


\section{Architecture}
\label{sec:architecture}

MemPool is an open-source shared-L1 many-core cluster with \num{256}
very-small cores sharing a multi-banked shared-L1 \gls{SPM} through a
low-latency interconnect~\cite{MemPool2021,MemPoolGitHub2021}. Its
maximum operating frequency shows a high sensibility to the available
routing resources and footprint size.

\subsection{Tile}
\label{sec:tile-micr}

\begin{figure}[h]
  \centering
  \includegraphics[width=.8\linewidth]{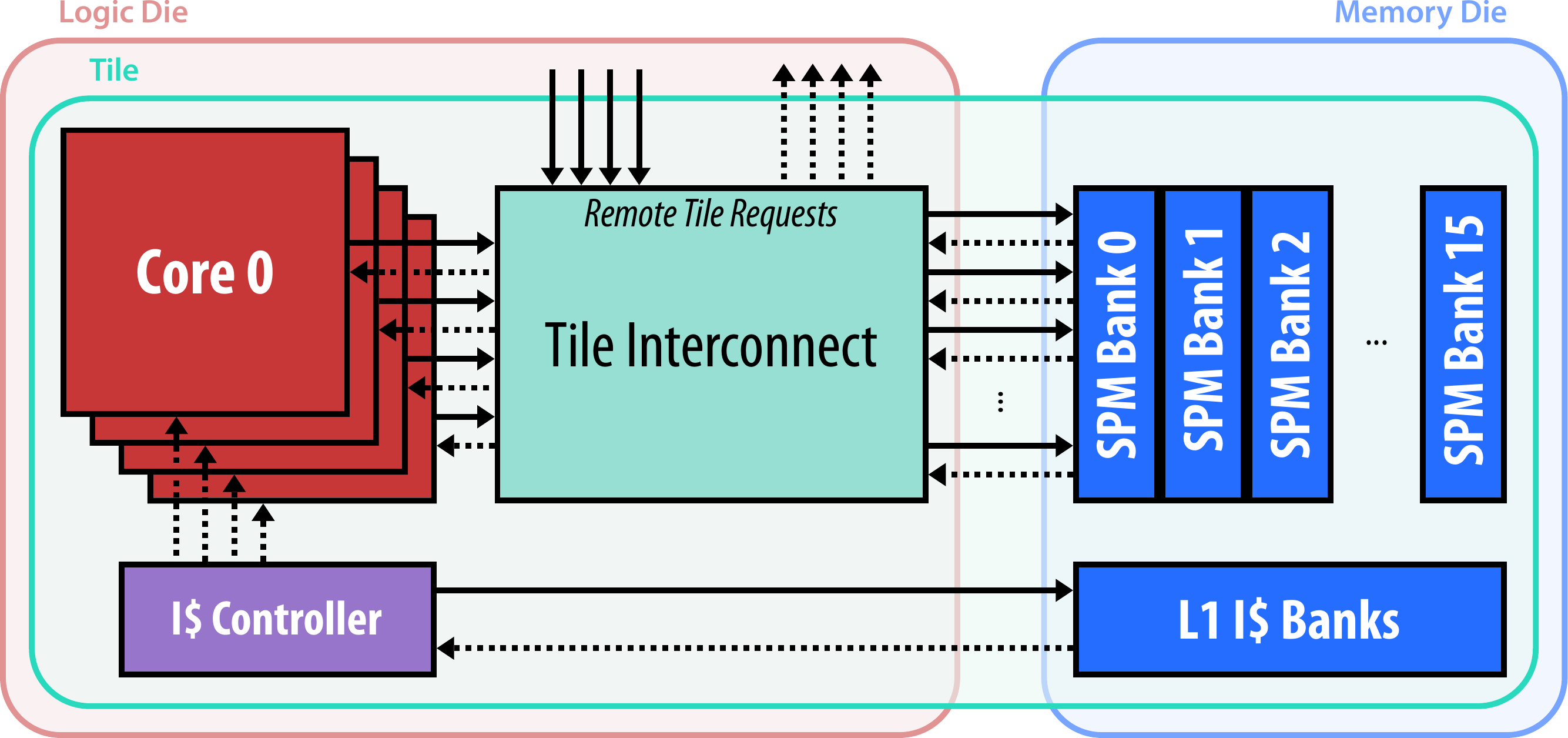}
  \caption{MemPool's tile architecture, highlighting a possible
    partitioning into the logic and the memory dies, used for the
    \SIlist{1;2;4}{\mebi\byte} 3D designs. Solid lines
    indicate requests, and dotted lines indicate responses.}
  \label{fig:tile_arch}
\end{figure}

MemPool is built hierarchically through the replication of tiles,
whose architecture can be seen in \Cref{fig:tile_arch}. Each tile
contains four very-small Snitch RV32IMAXpulpimg
cores~\cite{Zaruba2020}, \SI{2}{\kibi\byte} of L1 instruction cache,
and \num{16} \gls{SRAM} banks of \gls{SPM} locally accessible within
one cycle. The cores can execute instructions of the Xpulpimg
extension, \eg multiply-accumulate and load/store post-increment
instructions. A fully connected logarithmic crossbar connects local
cores and banks. In addition, four remote ports per tile allow remote
tiles to access the local tile's \gls{SPM} banks.

In our physical implementations with Macro-3D, the tile is partitioned
into a logic and a memory die. A possible partitioning would assign
all the memory banks of a tile to the memory die, namely
\SI{2}{\kibi\byte} of instruction cache and a multi-banked \gls{SPM}
of parameterizable capacity, as shown in \Cref{fig:tile_arch}. The
area requirement of \SI{60}{\kilo\gate} per Snitch
core~\cite{MemPool2021} and the tile interconnect's logic define the
footprint required for the logic die. With the default \gls{SPM}
capacity of \SI{1}{\mebi\byte} utilizing only \SI{51}{\percent} of the
memory die area, an increase of the \gls{SPM} capacity can be used to
balance the area requirement of both dies. As we will analyze in
\Cref{sec:tile-implementation}, the \SI{8}{\mebi\byte} 3D design uses
an adjusted partitioning scheme due to the increased \gls{SRAM} size.

\subsection{Group and Cluster}
\label{sec:group-cluster}

\begin{figure}[ht]
  \centering

  \begin{minipage}[t]{0.60\linewidth}
    \centering
    \includegraphics[width=.7\linewidth]{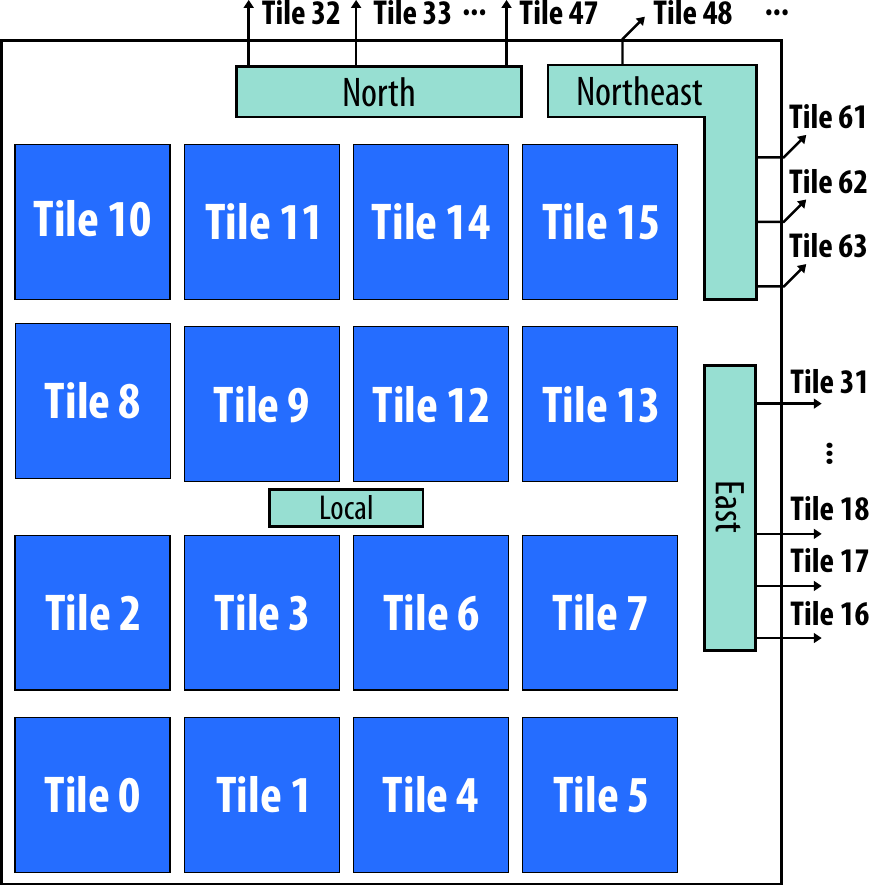}
    \subcaption{MemPool Group.}
    \label{fig:arch_group}
  \end{minipage}\hfill%
  \begin{minipage}[t]{0.35\linewidth}
    \centering
    \includegraphics[width=.9\linewidth]{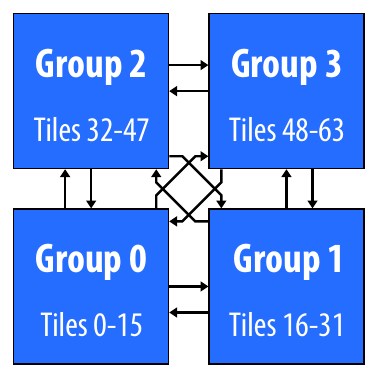}
    \subcaption{MemPool Cluster.}
    \label{fig:arch_cluster}
  \end{minipage}

  \caption{MemPool's hierarchical architecture, with \num{16} tiles
    forming a group (\Cref{fig:arch_group}) and \num{4} groups forming
    the full MemPool cluster (\Cref{fig:arch_cluster}).}
  \label{fig:arch_mempool}
\end{figure}

The MemPool cluster is built hierarchically using the tile as the
starting point. Sixteen tiles form a \emph{group}, whose architecture
is shown in \Cref{fig:arch_group}. Each core can access \gls{SPM}
banks in the same group within three cycles. Four \num{16x16} radix-4
butterfly networks are used in each group to connect tiles within the
same group (\emph{local} interconnect) or in different groups
(\emph{north}, \emph{northeast}, and \emph{east} interconnects).

At the top level of the architecture, the MemPool cluster has four
identical groups, as shown in \Cref{fig:arch_cluster}. Each core can
access \gls{SPM} banks in other groups with five cycles of latency. At
this hierarchical level, there are only point-to-point connections
between groups. This paper focuses on implementing the MemPool groups
since only a few cells (about five thousand) need to be placed between
them at the cluster level.

The dissimilarity between the tile and group makes MemPool a highly
suitable architecture for 3D integration. While the tile has a high
logic density, the group is highly congested due to its global
interconnects. The 2D MemPool's critical path goes from one tile to
the other diagonally opposed to it. Approximately \SI{37}{\percent} of
its timing is wire propagation delay, and \SI{75}{\percent} of its
cells are buffers or inverter pairs~\cite{MemPool2021}.


\section{Methodology}
\label{sec:methodology}

MemPool is synthesized and implemented in a commercial
\SI{28}{\nano\meter} high-$\kappa$ technology node. In the 3D
implementations, we use a fine \gls{F2F} via pitch of
\SI{1.0}{\micro\meter}. The F2F via size, resistance, and capacitance
is \SI{0.5x0.5}{\micro\meter}, \SI{0.5}{\ohm}, and
\SI{1}{\femto\farad} respectively~\cite{Beyne2017}. The 3D \gls{BEOL}
comprises six metal layers in both tiers (M6M6), separated by the
\gls{F2F} via layer, similarly to the setup proposed by
Bamberg~\etal\cite{Bamberg2020}.

Tiles are first synthesized using Synopsys Design Compiler 2021.06 and
then implemented with the corresponding flow in Cadence Innovus
20.13. We use a uniform \SI{1}{\giga\hertz} frequency target on the
typical corner to implement all the designs.  The 2D tiles use a
six-layer \gls{BEOL} (M6), while the 3D tiles use the mirrored M6M6
stack mentioned above. The implemented tiles---abstracted into black
boxes with full blockages on all utilized routing layers---are used for
the groups' syntheses and physical implementations, similarly to the
reference flow in \cite{MemPool2021}.

The 3D groups use the same \gls{BEOL} as the tiles, \ie M6M6, while
the 2D groups have two extra layers (M8) to allow over-the-tile
routing.  When abstracting a tile implementation, the physical
representation by default creates obstructions on all metal layers
that are available for routing. In the 2D case, this affects M1 to
M6. As the Macro-3D tile utilizes the \gls{BEOL} of the logic and the
memory die simultaneously, the tile abstraction blocks not only M1 to
M6 of the logic die, but also M1 to M6 of the memory die. The 3D tile
abstractions, therefore, prevent any inter-tile routing on the group
level where tiles are placed.

In this paper, we analyze a total of eight MemPool
configurations. Each configuration is named
MemPool-Flow$_{\text{Capacity}}$, where \emph{Capacity} is the total
capacity of the shared-L1 \gls{SPM} at the MemPool cluster, \ie one of
\SIlist{1;2;4;8}{\mebi\byte}, and \emph{Flow} is either 2D or 3D.



\section{Tile Implementation}
\label{sec:tile-implementation}

\begin{figure}[htbp]
  \centering
  \begin{minipage}[t]{0.33\linewidth}
    \centering
    \includegraphics[width=0.8235\linewidth]{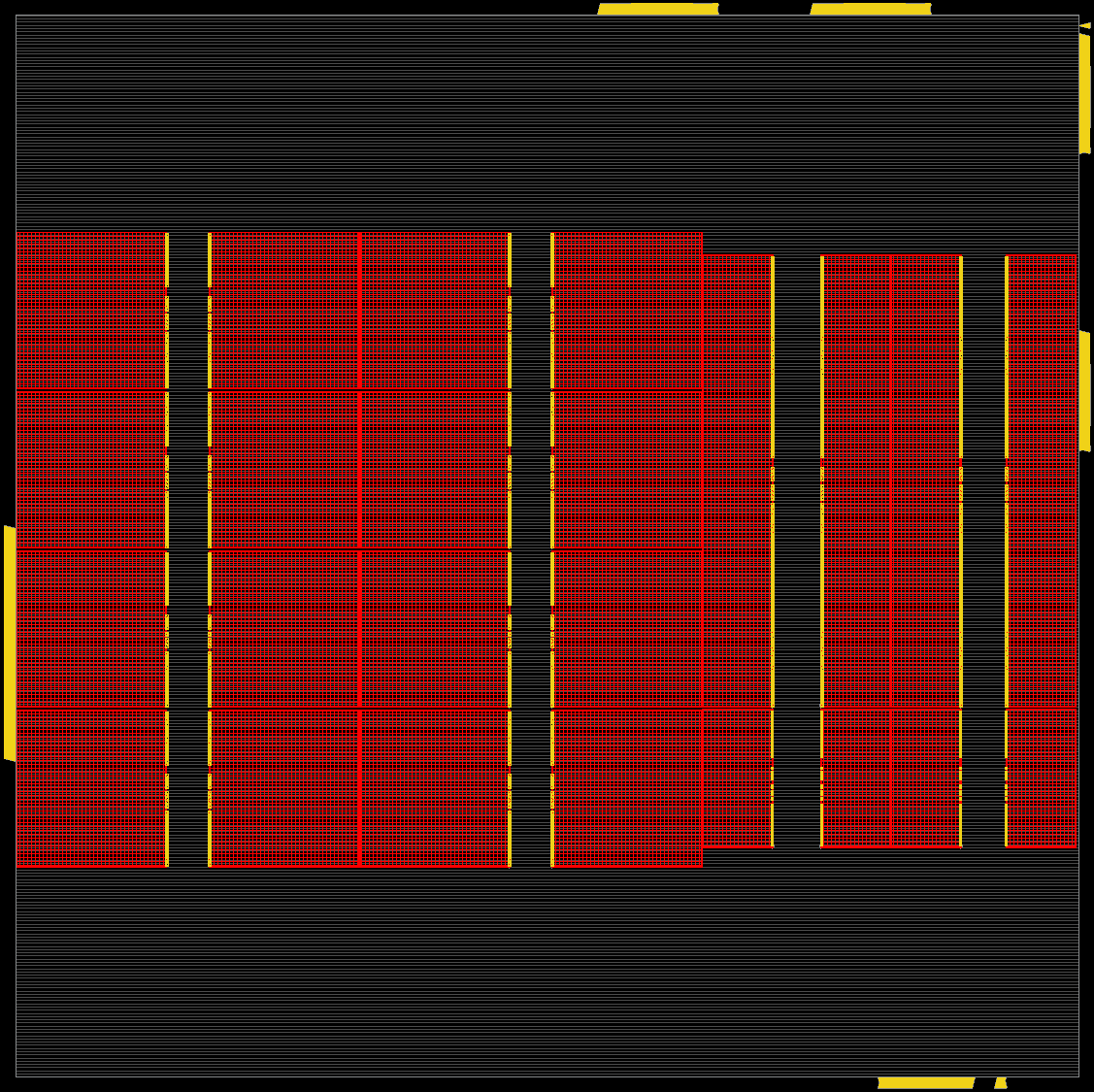}
    \subcaption{\mempool{1}{3D}.}
    \label{fig:tile_fp_1kib}
  \end{minipage}%
  \begin{minipage}[t]{0.33\linewidth}
    \centering
    \includegraphics[width=0.8739\linewidth]{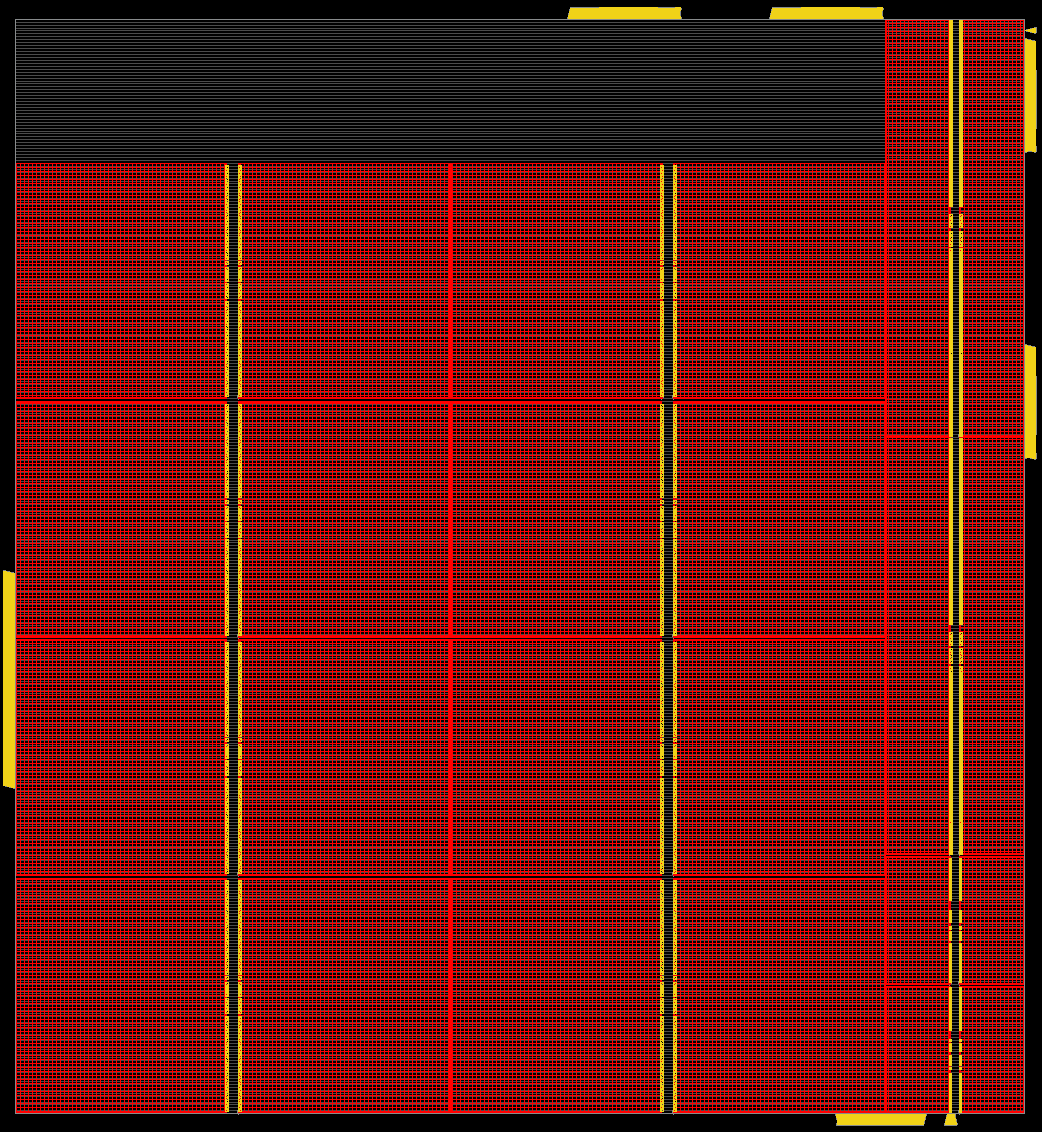}
    \subcaption{\mempool{4}{3D}.}
    \label{fig:tile_fp_4kib}
  \end{minipage}\hfill%
  \begin{minipage}[t]{0.33\linewidth}
    \centering
    \includegraphics[width=.9\linewidth]{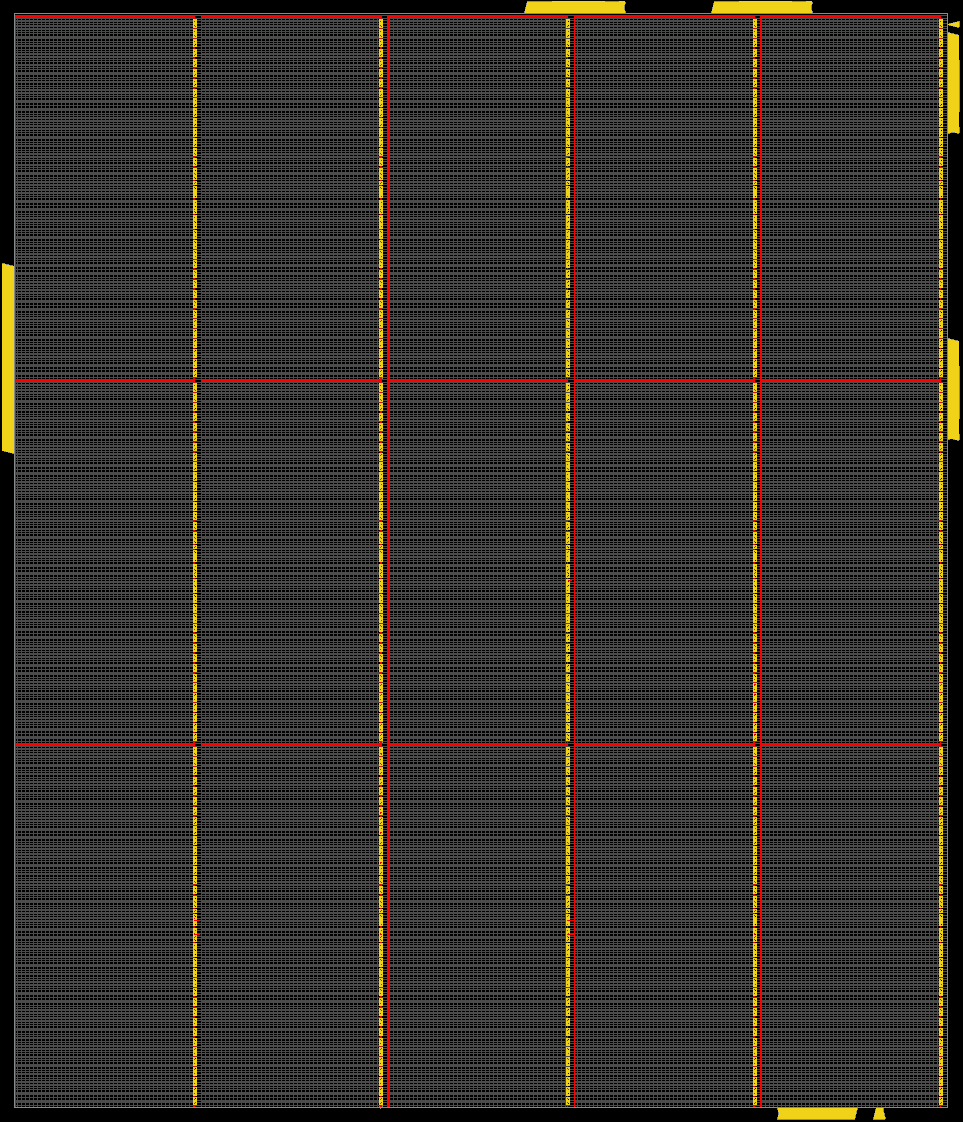}
    \subcaption{\mempool{8}{3D}.}
    \label{fig:tile_fp_8kib}
  \end{minipage}

  \caption{Floorplan of the memory die of the tile used to implement
    the \mempool{1}{3D}, \mempool{4}{3D}, and \mempool{8}{3D}. Images
    to scale.}
  \label{fig:tile_fp}
\end{figure}

The tiles are implemented to target a standard cell density of
\SI{90}{\percent} in the logic die. \Cref{fig:tile_fp} shows the
memory die floorplanning used to implement some of the considered
MemPool-3D configurations.  We use the partitioning of
\Cref{fig:tile_arch} to implement the tiles of the MemPool-3D
configurations with \SIrange{1}{4}{\mebi\byte} of \gls{SPM}. The
memory die of the \mempool{1}{3D} configuration, shown in
\Cref{fig:tile_fp_1kib}, only utilizes \SI{51}{\percent} of the area
of the memory die. On the other hand, the instance with
\SI{4}{\mebi\byte} of \gls{SPM}, shown in \Cref{fig:tile_fp_4kib},
achieves a much higher \SI{89}{\percent} utilization, with a tile
footprint only \SI{13}{\percent} higher than the \mempool{1}{3D} tile.

The \mempool{8}{3D} configuration uses a partitioning different from
the one shown in \Cref{fig:tile_arch} due to the increased area of the
memory macros. In this configuration, \num{15} out of the \num{16}
\gls{SPM} macros are arranged in a \num{5x3} array in the memory
die. This memory die achieves near \SI{100}{\percent} utilization, as
shown in \Cref{fig:tile_fp_8kib}, while its footprint is
\SI{40}{\percent} larger than the \mempool{1}{3D} design. As we can
see in \Cref{tab:tile_results}, both dies still keep a balanced area
ratio with this partitioning. The core utilization of the logic die is
\SI{84}{\percent}, only \SI{6}{\percent} below the target
utilization. The extra area of the logic die is used by one \gls{SPM}
bank and all the tile's instruction cache banks.

\begin{table}[htbp]
  \centering
  \caption{MemPool Tile's implementation results, normalized by the
    results of the baseline \mempool{1}{2D} configuration.}
  \begin{tabular}[h]{rrlll}
    \toprule
    \multirow{2}{*}{Flow} & \multirow{2}{*}{\gls{SPM} Capacity} & \multirow{2}{*}{Footprint} & \multicolumn{2}{c}{Core utilization}   \\\cmidrule{4-5}
                          &                                     &                            & \emph{Logic die}  & \emph{Memory die}  \\\midrule{}
    \multirow{4}{*}{2D}   & \SI{1}{\mebi\byte}                  & \num{1.000}                & \SI{90}{\percent} & ---                \\
                          & \SI{2}{\mebi\byte}                  & \num{1.104}                & \SI{90}{\percent} & ---                \\
                          & \SI{4}{\mebi\byte}                  & \num{1.420}                & \SI{84}{\percent} & ---                \\
                          & \SI{8}{\mebi\byte}                  & \num{1.817}                & \SI{86}{\percent} & ---                \\\midrule{}
    \multirow{4}{*}{3D}   & \SI{1}{\mebi\byte}                  & \num{0.667}                & \SI{90}{\percent} & \SI{51}{\percent}  \\
                          & \SI{2}{\mebi\byte}                  & \num{0.667}                & \SI{90}{\percent} & \SI{65}{\percent}  \\
                          & \SI{4}{\mebi\byte}                  & \num{0.767}                & \SI{85}{\percent} & \SI{89}{\percent}  \\
                          & \SI{8}{\mebi\byte}                  & \num{0.933}                & \SI{84}{\percent} & \SI{100}{\percent} \\\bottomrule{}
  \end{tabular}
  \label{tab:tile_results}
\end{table}

There is a negligible \gls{PPA} difference across all tile
instances. The fastest tile, \mempool{4}{3D}, achieves a
frequency only \SI{6}{\percent} higher than the slowest tile,
\mempool{2}{3D}. This is because the tile is primarily constrained by
external delays that model the group rather than internal
register-to-register paths. While 3D integration can reduce the wire
length of these paths, the primary effect is reducing the tile's
footprint, leading to shorter interconnects and, thus, group level
\gls{PPA} improvements.


\section{Group Implementation}
\label{sec:group-impl-results}

\subsection{Area and Footprint}
\label{sec:footprint}

\begin{figure}[htbp]
  \centering
  \begin{minipage}[t]{0.48\linewidth}
    \centering
    \includegraphics[width=0.75\linewidth]{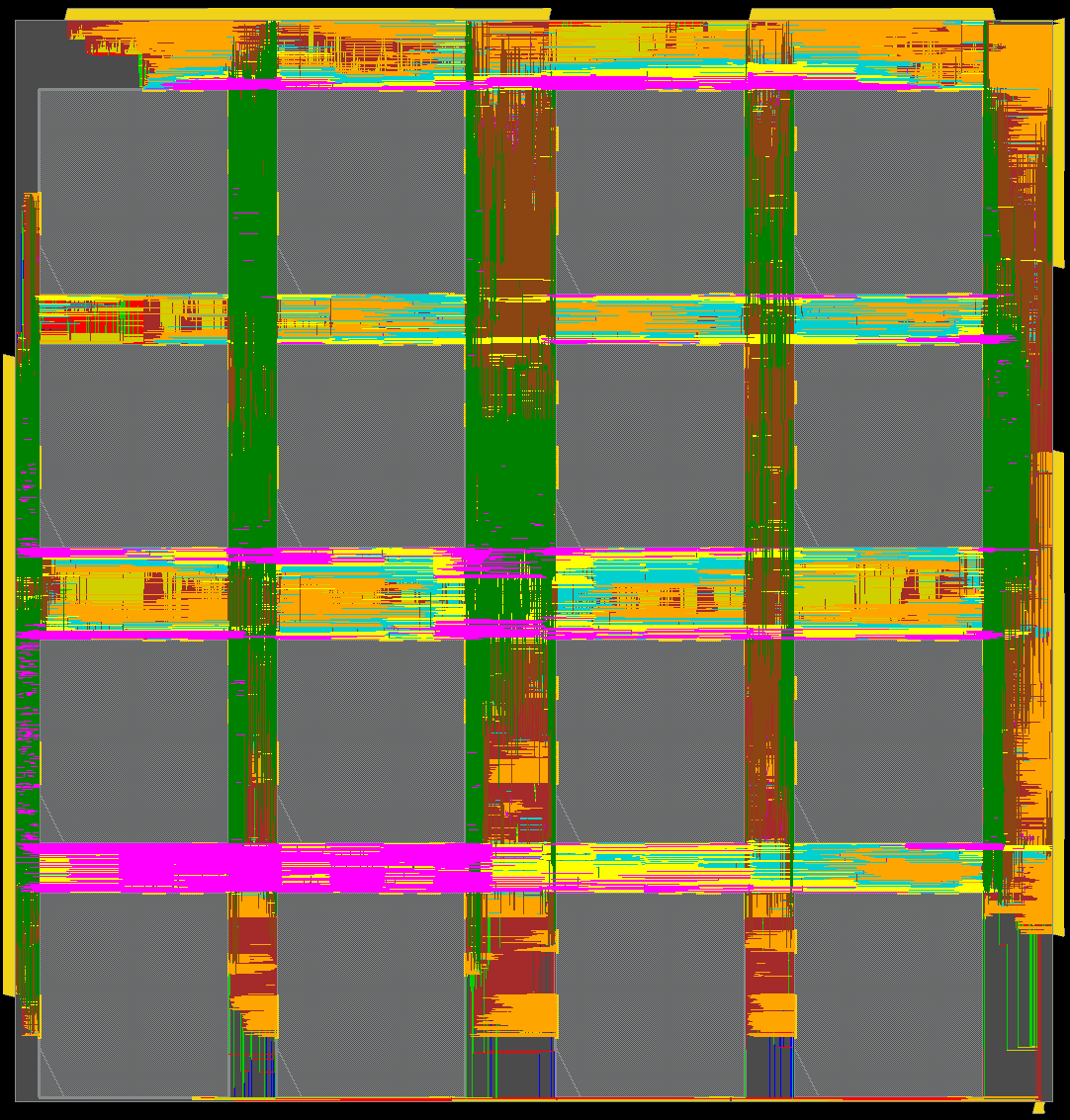}
    \subcaption{Routing.}
    \label{fig:routing_group_3d_4kib}
  \end{minipage}\hfill%
  \begin{minipage}[t]{0.48\linewidth}
    \centering
    \includegraphics[width=0.75\linewidth]{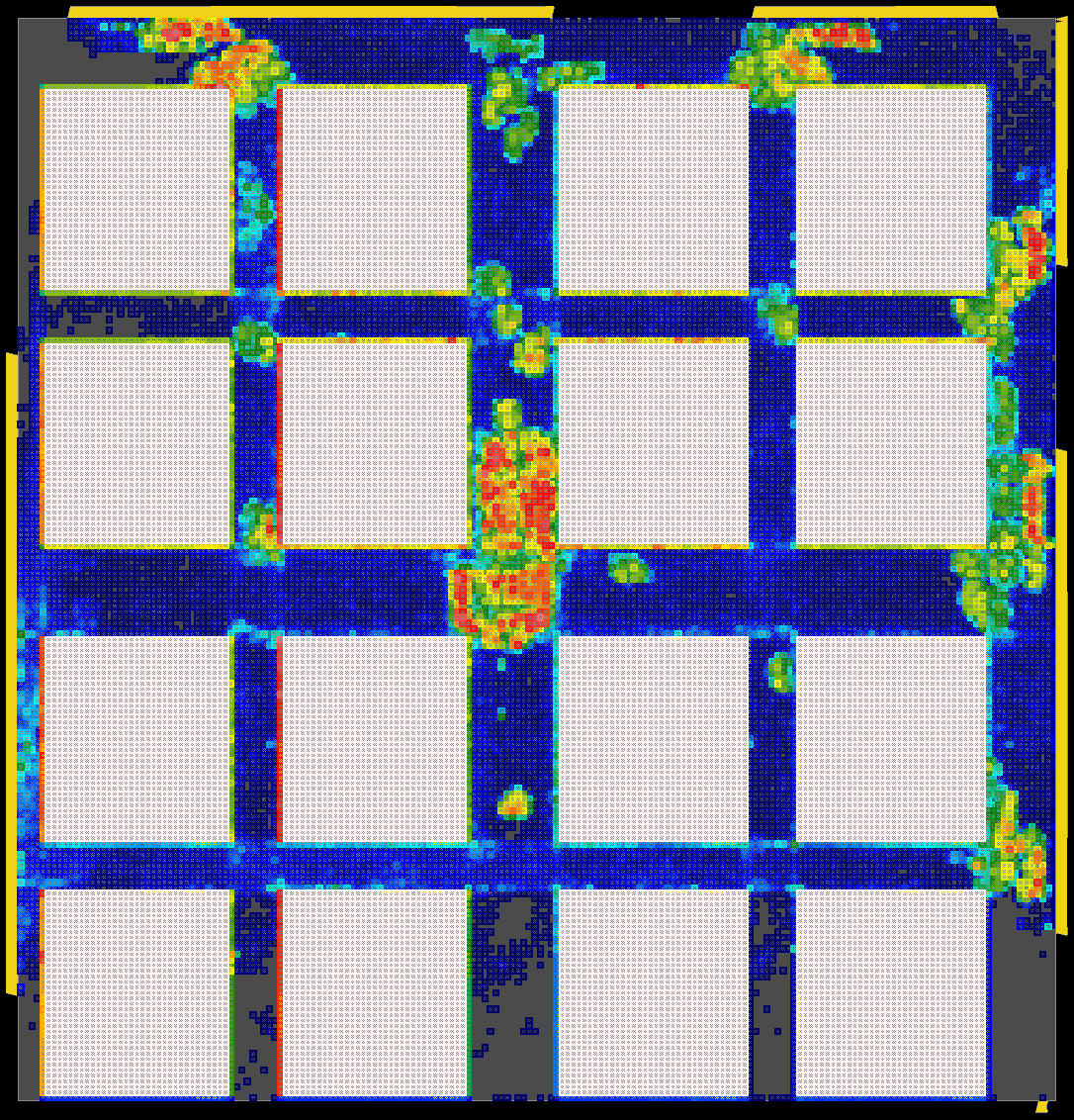}
    \subcaption{Cell density map.}
    \label{fig:cell_density_group_3d_4kib}
  \end{minipage}
  \caption{Routing and cell density map of the \mempool{4}{3D}
    group. Yellow and red colors indicate regions with a very high
    cell density. Dark blue and gray regions have a cell density close
    to zero.}
  \label{fig:routing_cell_density_group_3d_4kib}
\end{figure}

The group is at the critical level in the implementation of
MemPool. When deriving the channel widths between tiles, we need to
consider that the group is densely connected at the design's center,
where most of the logic of the local interconnect is placed. This
causes heavy congestion, creating \glspl{DRV} and degrades timing if
the tiles are not sufficiently spaced in the center of the
design. \Cref{fig:routing_cell_density_group_3d_4kib} exemplifies this
by showing the routing and cell density map of the \mempool{4}{3D}
instance. The four group interconnects can be seen as pockets of very
high cell density in \Cref{fig:cell_density_group_3d_4kib}.

\begin{figure}[htbp]
  \centering
  \begin{minipage}[t]{0.48\linewidth}
    \centering
    \includegraphics[width=0.8\linewidth]{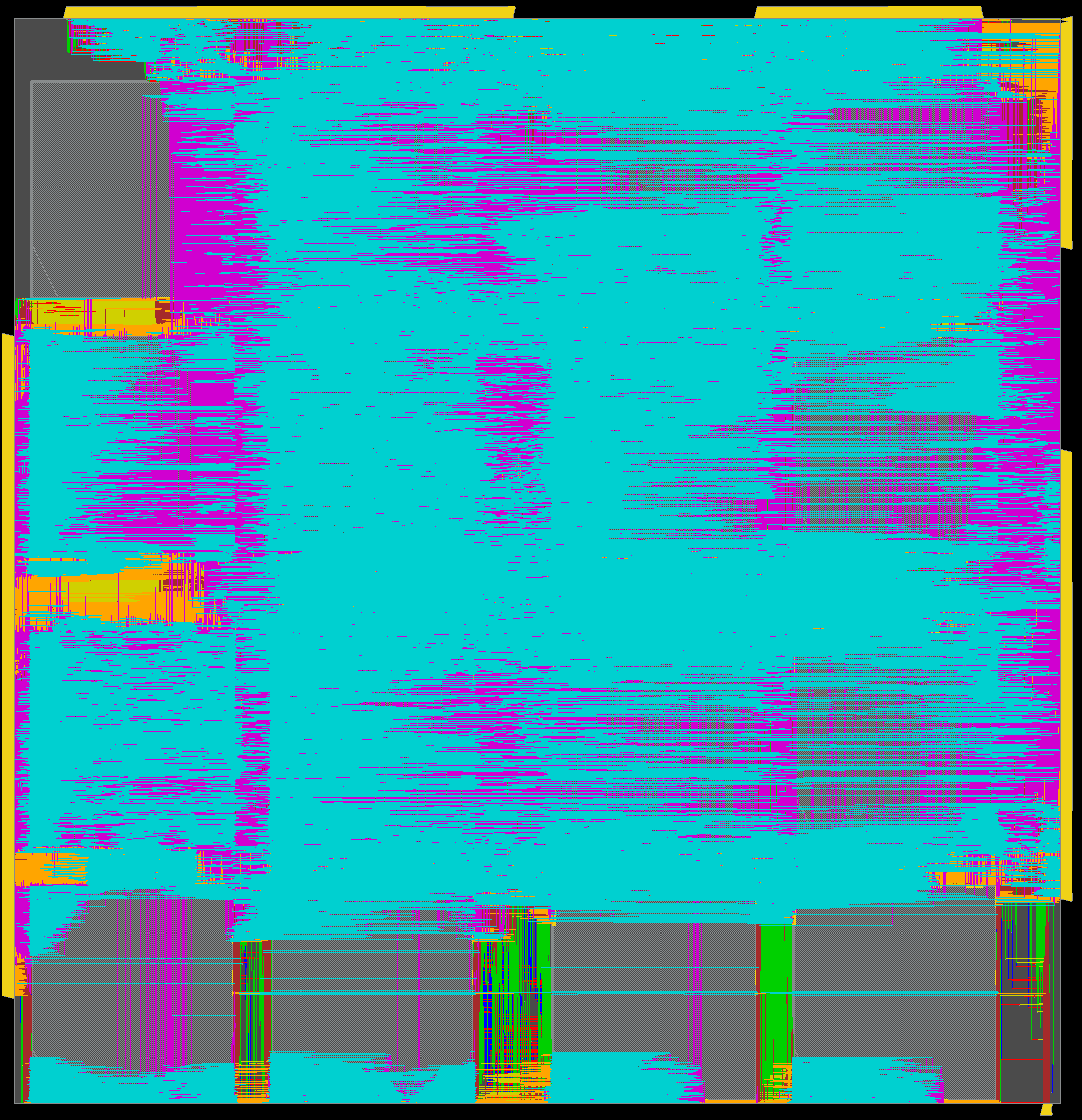}
    \subcaption{\mempool{8}{2D}.}
    \label{fig:group_2d_8kib}
  \end{minipage}\hfill%
  \begin{minipage}[t]{0.48\linewidth}
    \centering
    \includegraphics[width=0.6\linewidth]{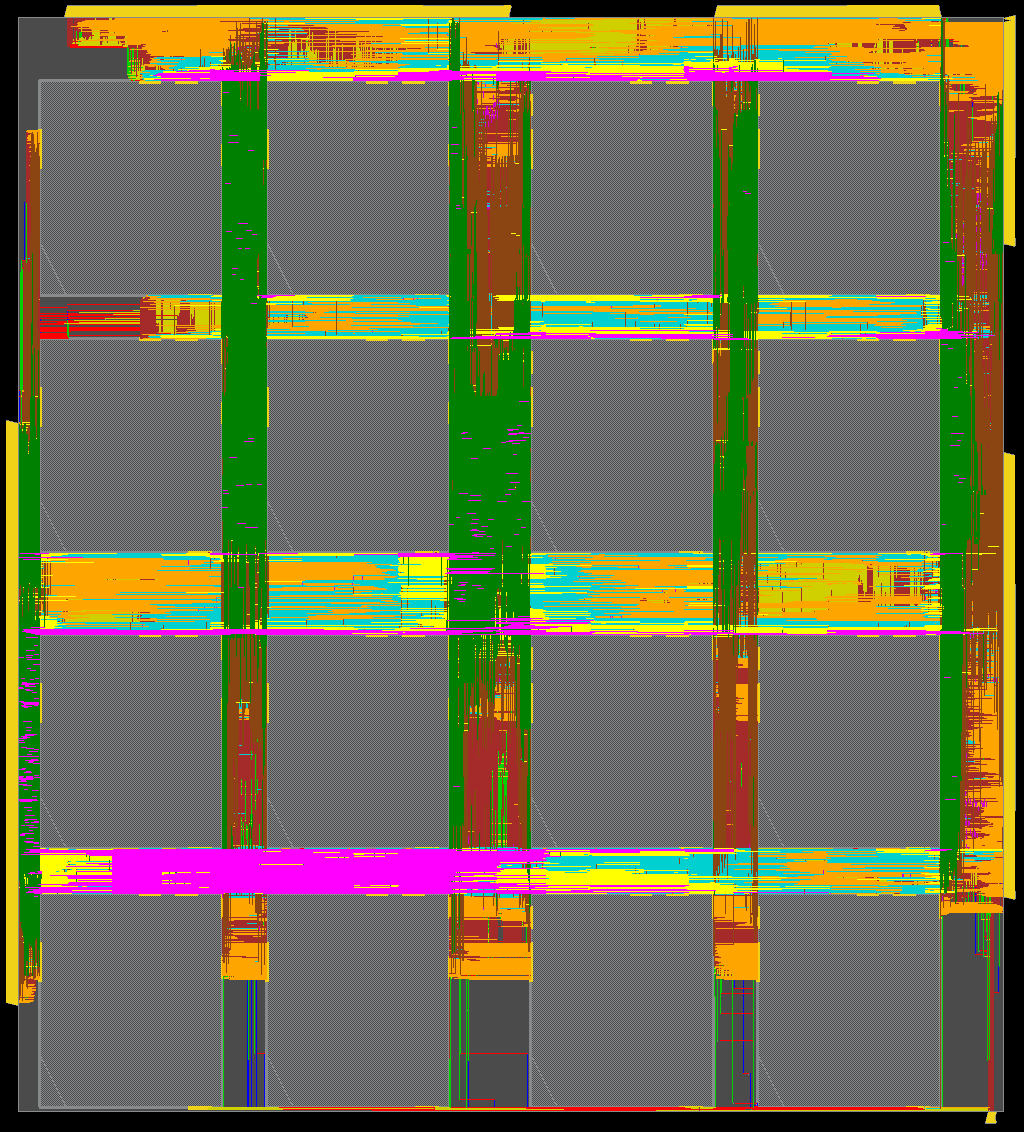}
    \subcaption{\mempool{8}{3D}.}
    \label{fig:group_3d_8kib}
  \end{minipage}
  \caption{\mempool{8}{2D} and \mempool{8}{3D} placed-and-routed
    designs, highlighting the group routing. Images to scale.}
  \label{fig:group_fp}
\end{figure}

\Cref{fig:group_fp} shows both group designs with \SI{8}{\mebi\byte}
of \gls{SPM}.  We highlight that the width of the channels between
tiles is kept constant for all trials of each flow. The rationale is
that the group interconnects' size is largely independent of the
\gls{SPM} capacity, except for the additional address bits.

The channels between the 3D tiles are \SI{18}{\percent} narrower than
the 2D counterparts since the 3D designs have the twelve layers of the
mirrored M6M6 \gls{BEOL} to route the group interconnects. In
contrast, the 2D trials can only use the eight layers of the M8
\gls{BEOL}. This effect compounds with the reduced footprint of the 3D
tiles, further reducing the footprint of the MemPool-3D groups.  In
\Cref{fig:group_fp}, we can also see the over-the-tile routing of the
2D designs---since the tiles are routed up to M6 and the group up to
M8. Even though the 3D runs have a more aggressive \gls{BEOL} than the
2D runs, M6M6, the lack of over-the-tile routing incurs extra
congestion since all group interconnects have routing confined to the
channels. The footprint of the MemPool-3D groups could be further
reduced if routing resources of the tiles would be available to
implement the group level.

\Cref{tab:results} summarizes the implementation results of the eight
considered groups, normalized by the baseline group
\mempool{1}{2D}. In terms of footprint, the 3D groups are much smaller
than the 2D groups. For example, the largest 3D group,
\mempool{8}{3D}, has a footprint \SI{14}{\percent} smaller than the
smallest 2D group, \mempool{1}{2D}. The 3D MemPool groups also have a
footprint much smaller than their 2D counterparts. For instance, the
\mempool{8}{3D} group has a footprint \SI{46}{\percent} smaller than
\mempool{8}{2D}, as seen in \Cref{fig:group_fp}.

\begin{table*}[t]
  \centering
  \caption{MemPool group's 2D and 3D power, operating frequency,
    footprint, area, and energy efficiency results of MemPool
    instances with \SIlist{1;2;4;8}{\mebi\byte} of \gls{SPM} L1
    memory, normalized by the results of the baseline 2D design with
    \SI{1}{\mebi\byte} of \gls{SPM}. The numbers in parentheses show
    the difference relative to the 2D counterpart design.}
  \begin{tabular}[h]{rllcllcllcll}
    \toprule
    L1 \gls{SPM} Capacity   & \multicolumn{2}{c}{\SI{1}{\mebi\byte}}            &  & \multicolumn{2}{c}{\SI{2}{\mebi\byte}}            &  & \multicolumn{2}{c}{\SI{4}{\mebi\byte}}            &  & \multicolumn{2}{c}{\SI{8}{\mebi\byte}}            \\\cmidrule{2-3}\cmidrule{5-6}\cmidrule{8-9}\cmidrule{11-12}
    Implementation Flow     & 2D            & 3D                                &  & 2D            & 3D                                &  & 2D            & 3D                                &  & 2D            & 3D                                \\\midrule{}
    \gls{BEOL}              & M8            & M6M6                              &  & M8            & M6M6                              &  & M8            & M6M6                              &  & M8            & M6M6                              \\
    Footprint               & \num{1.000}   & \num{0.665} (\SI{-33}{\percent})  &  & \num{1.074}   & \num{0.665} (\SI{-38}{\percent})  &  & \num{1.299}   & \num{0.737} (\SI{-43}{\percent})  &  & \num{1.572}   & \num{0.857} (\SI{-46}{\percent})  \\
    Combined Die Area       & \num{1.000}   & \num{1.330} (\SI{+33}{\percent})  &  & \num{1.074}   & \num{1.330} (\SI{+23}{\percent})  &  & \num{1.299}   & \num{1.474} (\SI{+13}{\percent})  &  & \num{1.572}   & \num{1.714} (\SI{+9.0}{\percent}) \\
    Wire Length             & \num{1.000}   & \num{0.803}                       &  & \num{1.036}   & \num{0.803}                       &  & \num{1.131}   & \num{0.844}                       &  & \num{1.294}   & \num{0.888}                       \\
    Density [\si{\percent}] & \num{53.0}    & \num{54.5}                        &  & \num{54.0}    & \num{54.8}                        &  & \num{53.4}    & \num{53.2}                        &  & \num{56.9}    & \num{54.4}                        \\\midrule{}
    \#Buffers               & \num{182.9e3} & \num{151.5e3}                     &  & \num{190.3e3} & \num{151.2e3}                     &  & \num{212.5e3} & \num{166.5e3}                     &  & \num{217.6e3} & \num{156.1e3}                     \\
    \#\gls{F2F} Bumps       & ---           & \num{78.3e3}                      &  & ---           & \num{78.9e3}                      &  & ---           & \num{84.4e3}                      &  & ---           & \num{86.2e3}                      \\\midrule{}
    Eff. Frequency          & \num{1.000}   & \num{1.040} (\SI{+4.0}{\percent}) &  & \num{0.930}   & \num{0.979} (\SI{+5.2}{\percent}) &  & \num{0.875}   & \num{0.955} (\SI{+9.1}{\percent}) &  & \num{0.885}   & \num{0.930} (\SI{+5.1}{\percent}) \\
    Total Negative Slack    & \num{-1.000}  & \num{-0.184}                      &  & \num{-2.080}  & \num{-0.458}                      &  & \num{-5.887}  & \num{-0.604}                      &  & \num{-5.212}  & \num{-0.962}                      \\
    \#Failing Paths         & \num{1140}    & \num{1046}                        &  & \num{1636}    & \num{1332}                        &  & \num{4396}    & \num{1747}                        &  & \num{4352}    & \num{2403}                        \\\midrule{}
    Total Power             & \num{1.000}   & \num{0.913}                       &  & \num{1.045}   & \num{0.958}                       &  & \num{1.129}   & \num{1.041}                       &  & \num{1.299}   & \num{1.173}                       \\
    Power-delay product     & \num{1.000}   & \num{0.877} (\SI{-12}{\percent})  &  & \num{1.129}   & \num{0.981} (\SI{-13}{\percent})  &  & \num{1.290}   & \num{1.089} (\SI{-16}{\percent})  &  & \num{1.469}   & \num{1.261} (\SI{-14}{\percent})  \\\bottomrule
  \end{tabular}
  \label{tab:results}
\end{table*}

The combined area of the memory and logic dies of the MemPool-3D
groups is larger than the area of the MemPool-2D groups. The area
overhead of the MemPool-3D groups, however, decreases with increasing
\gls{SPM} capacity. The combined die area of the largest 3D group,
\mempool{8}{3D}, is only \SI{9}{\percent} larger than the
\mempool{8}{2D} group area, indicating that the partitioning is closer
to ideal. Although the footprint is the most important metric for
analyzing \gls{PPA} gains of the 3D integration thanks to reduced
interconnect lengths, the combined area is more relevant for an
implementation cost analysis of the 3D designs. This paper focuses on
the physical implementation of the groups, MemPool's most critical
hierarchical level since the cluster only has four identical groups
and some glue logic. However, it can be noted that the 12-layer
mirrored \gls{BEOL} of the MemPool-3D designs implies that the
channels between groups needed to route the cluster-level connections
can be made shorter than the equivalent channels of the MemPool-2D
cluster. This means that we can expect an even more favorable area
ratio at the cluster level.

\subsection{Power and Operating Frequency}
\label{sec:power-oper-freq}

MemPool's sensibility to the footprint size can be seen in the
normalized wire length results of \Cref{tab:results}. The wire length
of the MemPool-2D groups grows by \SI{29.4}{\percent} from
\mempool{1}{2D} to \mempool{8}{2D}, accompanied by an
\SI{18.9}{\percent} increase in the number of buffers. The effect
compounds to other \gls{PPA} metrics. For example, the 2D groups
achieve an operating frequency up to \SI{12.5}{\percent} slower, a
power consumption up to \SI{29.9}{\percent} higher, and a \gls{PDP} up
to \SI{46.9}{\percent} higher than the baseline \mempool{1}{2D} group.

The MemPool-3D groups have a significantly smaller \gls{PPA}
degradation with increasing \gls{SPM} capacity than the MemPool-2D
groups. The low utilization of the \mempool{1}{3D} design
(\Cref{fig:tile_fp_1kib}) implies it is possible to implement the
\mempool{2}{3D} group without increasing the footprint. Even the
largest MemPool-3D group, with \SI{8}{\mebi\byte} of \gls{SPM}, has a
footprint only \SI{10.6}{\percent} larger than the \mempool{1}{3D}
group. This smaller footprint variation leads to a \gls{PPA}
degradation which is less drastic than the one affecting the
MemPool-2D groups. As a result, the MemPool-3D groups achieve an
operating frequency only \SI{11.8}{\percent} slower and a power
consumption only \SI{28.4}{\percent} higher than the baseline
\mempool{1}{3D} group.

The benefits of 3D integration on MemPool are clearer when comparing
instances with the same \gls{SPM} capacity across 2D and 3D
implementation flows. In general, the MemPool-3D designs have a
smaller footprint, achieve a higher operating frequency, and consume
less power than their 2D counterparts. In terms of footprint, the
largest gains are found on the groups of the \SI{8}{\mebi\byte}
configuration, with \mempool{8}{3D} having a footprint
\SI{46}{\percent} smaller than \mempool{8}{2D}. In terms of frequency
and \gls{PDP}, the \mempool{4}{3D} group achieves an operating
frequency \SI{9.1}{\percent} higher and a \gls{PDP} \SI{16}{\percent}
lower than the \mempool{4}{2D} group. It is also interesting that
there is an operating frequency drop of \SI{6.2}{\percent} between the
\mempool{2}{3D} and \mempool{1}{3D} groups, despite having the same
footprint. This is due to the longer \glspl{SRAM}' delay, which
impacts the timing of the tile's input-to-register and
register-to-output paths. In general, the MemPool-3D groups have a
higher operating frequency, consume less power, and have a lower
\gls{PDP} than the MemPool-2D groups with the same \gls{SPM} capacity
thanks to their smaller footprint and wire length.


\section{Performance Analysis}
\label{sec:mempools-performance}

We use a matrix multiplication kernel as a representative application
for MemPool's target domain to quantify the algorithmic benefits of
increasing the memory capacity even for compute-bound
kernels. Benefits on memory bound kernels are obviously larger, but we
believe the analysis in a compute-bound regime allows us to gather
more interesting insights on MemPool-3D's performance.

\subsection{Cycle count}
\label{sec:cycle-count}

We measure the cycle count of computing the matrix multiplication of
two $M \times M$ ($M=326400$) matrices that do not fit into the \gls{SPM} but
reside in global memory through a cycle-accurate register-transfer level simulation of the \mempool{} cluster. The matrix
size is chosen to be the least common multiple of the tile sizes
$t \times t$ ($t=256,384,544,800$) that fully utilize the available \gls{SPM} in each
configuration, enabling optimal tiling and maximizing data reuse.  The cores load the input
tiles and synchronize in a memory phase before computing on the output
tile in a compute phase. Those phases are repeated until the output
tile is fully computed. At this point, the output tile is stored back
into the main memory, and the process repeats for the subsequent
output tiles. Since different output tiles require the same input
data, each input element is loaded exactly $M/t$ times. Therefore,
having a bigger \gls{SPM} allows for more data reuse and less memory
overhead. The second benefit of increased tile size is the increased
length of the compute phase minimizing repeated static overhead due to
loop setup and synchronization.

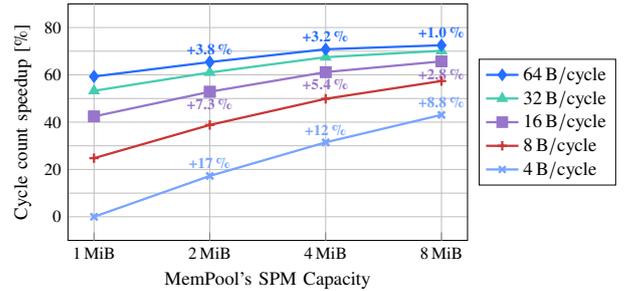
\begin{figure}[htbp]
  \centering
  \resizebox{.9\linewidth}{!}{
  \begin{tikzpicture}
    \begin{axis}[
      height=6cm,
      width=\linewidth,
      enlargelimits=0,
      scaled y ticks = false,
      legend style={
        at={(1.2,0.5)},
        anchor=center,
        legend columns=1
      },
      legend cell align={left},
      ylabel={Cycle count speedup [\si{\percent}]},
      ymajorgrids=true,
      yminorgrids=true,
      ymin=-10,
      ymax=90,
      minor y tick num=1,
      ytick style={draw=none},
      xlabel={MemPool's \gls{SPM} Capacity},
      xmode=log,
      log basis x={2},
      log ticks with fixed point,
      xtick=\empty,
      xmajorgrids=true,
      extra x ticks={1, 2, 4, 8},
      extra x tick labels={\SI{1}{\mebi\byte}, \SI{2}{\mebi\byte}, \SI{4}{\mebi\byte}, \SI{8}{\mebi\byte}},
      xtick pos=left,
      enlarge x limits=0.075]
      \pgfplotstableread{fig/benchmarks/matmul_speedup_relative.csv}\loadedtable;
      \addplot [color=color1, very thick, mark=diamond*, mark size=2.5pt] table[x=Memory, y expr=100*\thisrowno{5}] {\loadedtable};
      \addplot [color=color2, very thick, mark=triangle*, mark size=2.5pt] table[x=Memory, y expr=100*\thisrowno{4}] {\loadedtable};
      \addplot [color=color3, very thick, mark=square*, mark size=2.5pt] table[x=Memory, y expr=100*\thisrowno{3}] {\loadedtable};
      \addplot [color=color4, very thick, mark=+, mark size=2.5pt] table[x=Memory, y expr=100*\thisrowno{2}] {\loadedtable};
      \addplot [color=color5, very thick, mark=x, mark size=2.5pt] table[x=Memory, y expr=100*\thisrowno{1}] {\loadedtable};

      \addplot [color=color1, very thick, mark=diamond*, mark size=2.5pt, only marks] table[x=Memory, y expr=100*\thisrowno{5}] {\loadedtable}
      node [above, pos=0.33](64B_2MiB){\footnotesize\textbf{\SI{+3.8}{\percent}}}
      node [above, pos=0.67](64B_4MiB){\footnotesize\textbf{\SI{+3.2}{\percent}}}
      node [above, pos=1.00](64B_8MiB){\footnotesize\textbf{\SI{+1.0}{\percent}}};
      \addplot [color=color3, very thick, mark=square*, mark size=2.5pt, only marks] table[x=Memory, y expr=100*\thisrowno{3}] {\loadedtable}
      node [below, pos=0.33](16B_2MiB){\footnotesize\textbf{\SI{+7.3}{\percent}}}
      node [below, pos=0.67](16B_4MiB){\footnotesize\textbf{\SI{+5.4}{\percent}}}
      node [below, pos=1.00](16B_8MiB){\footnotesize\textbf{\SI{+2.8}{\percent}}};
      \addplot [color=color5, very thick, mark=x, mark size=2.5pt, only marks] table[x=Memory, y expr=100*\thisrowno{1}] {\loadedtable}
      node [above, pos=0.33](4B_2MiB){\footnotesize\textbf{\SI{+17}{\percent}}}
      node [above, pos=0.67](4B_4MiB){\footnotesize\textbf{\SI{+12}{\percent}}}
      node [above, pos=1.00](4B_8MiB){\footnotesize\textbf{\SI{+8.8}{\percent}}};

      \legend{
        \SI{64}{\byte\per\cycle},
        \SI{32}{\byte\per\cycle},
        \SI{16}{\byte\per\cycle},
        \SI{8}{\byte\per\cycle},
        \SI{4}{\byte\per\cycle}};
    \end{axis}
  \end{tikzpicture}}
\caption{Cycle count speedup of the matrix multiplication kernel with
  larger \gls{SPM} capacity, as a function of the off-chip memory
  bandwidth, relative to the \SI{1}{\mebi\byte} configuration with an
  off-chip memory bandwidth of \SI{4}{\byte\per\cycle}. The
  percentages by the data points indicate the speedup relative to the
  instance with the same off-chip memory bandwidth but half of the
  \gls{SPM} capacity.}
  \label{fig:speedup_analysis}
\end{figure}

We calculate the cycle count of the memory phase for different
off-chip memory bandwidths. A classic \gls{DDR} \gls{SDRAM} stick has
a data width of \SI{8}{\byte}, which means a single \gls{DDR} channel
clocked at the same frequency as MemPool could deliver at most
\SI{16}{\byte\per\cycle}. Therefore, we analyze
bandwidths around this realistic case, ranging from a worst-case
bandwidth of \SI{4}{\byte\per\cycle} to a very optimistic
\SI{64}{\byte\per\cycle}.  Our model idealizes the latency into the
off-chip global memory. We measure the duration of the compute phase
with a hot instruction cache and calculate the total cycle count by
accumulating all phases. The results in \Cref{fig:speedup_analysis}
show a speedup of \SI{43} {\percent} for the \SI{8}{\mebi\byte} case
over the baseline for the worst-case bandwidth, where the memory
transfers make up a significant portion of the runtime. For the
off-chip memory bandwidth of one \gls{DDR} channel, the configuration
with \SI{8}{\mebi\byte} of \gls{SPM} achieves a cycle count speedup of
\SI{16}{\percent} over the baseline. Even for the optimistic off-chip
memory bandwidth of \SI{64}{\byte\per\cycle}, the largest
configuration still generates an \SI{8}{\percent} benefit over the
baseline, showing that a higher memory capacity is highly beneficial
from an algorithmic point of view.

\subsection{Performance and Energy-Efficiency}
\label{sec:physical-results}

The increased \gls{SPM} capacity impacts the \gls{PPA} of the design,
as seen in Section~\ref{sec:group-impl-results}. This Section combines
those effects to analyze how MemPool's performance and energy
efficiency evolve across all considered configurations. Throughout
this section, we chose an off-chip memory bandwidth of
\SI{16}{\byte\per\cycle} to represent our memory subsystem.

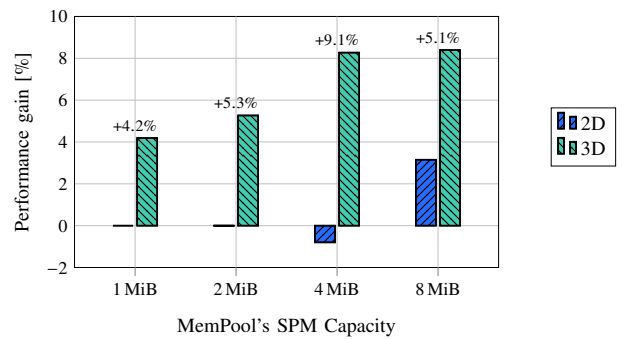
\begin{figure}[htpb]
  \centering
  \resizebox{.9\linewidth}{!}{
  \begin{tikzpicture}
    \begin{axis}[
      ybar,
      height=6cm,
      width=\linewidth,
      enlargelimits=0,
      scaled y ticks = false,
      legend style={
        at={(1.2,0.52)},
        anchor=center,
        legend columns=1
      },
      legend cell align={left},
      ylabel={Performance gain [\si{\percent}]},
      ymajorgrids=true,
      yminorgrids=true,
      ytick={-2,0,2,4,6,8,10},
      minor y tick num=0,
      ytick style={draw=none},
      ymin=-2,
      ymax=10,
      xlabel={MemPool's \gls{SPM} Capacity},
      xmode=log,
      log basis x={2},
      log ticks with fixed point,
      xmajorgrids=true,
      xtick pos=left,
      xtick=\empty,
      extra x ticks={1, 2, 4, 8},
      extra x tick labels={\SI{1}{\mebi\byte}, \SI{2}{\mebi\byte}, \SI{4}{\mebi\byte}, \SI{8}{\mebi\byte}},
      enlarge x limits=0.2]

      \pgfplotstableread{fig/benchmarks/matmul_performance.csv}\loadedtable;
      \addplot [fill=color1, thick, postaction={pattern=north east lines}] table [x=Memory, y expr=100*\thisrowno{1}-100] {\loadedtable};
      \addplot [fill=color2, thick, postaction={pattern=north west lines}] table[x=Memory, y expr=100*\thisrowno{2}-100] {\loadedtable};

      \addplot [only marks,
                point meta=explicit,
                nodes near coords={\footnotesize\pgfmathprintnumber[precision=1,showpos]{\pgfplotspointmeta}\%},
                nodes near coords style={above}]
         table [x=Memory,y expr={max(100*\thisrowno{1}-100,100*\thisrowno{2}-100)},
                meta expr={(\thisrowno{2}-\thisrowno{1})/\thisrowno{1} * 100}] {\loadedtable};

      \legend{
        2D,
        3D};
    \end{axis}
  \end{tikzpicture}}
\caption{Performance gain of the matrix multiplication kernel with
  larger \gls{SPM} capacity, relative to \mempool{1}{2D}
  with a \SI{16}{\byte\per\cycle} off-chip memory
  bandwidth. The percentages above the bars indicate the speedup of
  the MemPool-3D instance compared to the MemPool-2D with the same
  \gls{SPM} capacity.}
  \label{fig:performance_analysis}
\end{figure}

\Cref{fig:performance_analysis} shows the performance of the matrix
multiplication kernel on MemPool, as a function of the \gls{SPM}
capacity, for the MemPool-2D and MemPool-3D designs. Thanks to their
higher operating frequencies, the MemPool-3D groups achieve a
performance up to \SI{9.1}{\percent} higher than the MemPool-2D
groups. The MemPool-2D groups achieve small performance gains with
increasing \gls{SPM} capacity, reaching a gain of at most
\SI{3.1}{\percent} for the \mempool{8}{2D} case. Due to a particularly
low operating frequency, the \mempool{4}{2D} has a performance drop
compared to the \mempool{1}{2D} design. The MemPool-3D designs, on the
other hand, achieve consistently higher performances with increasing
\gls{SPM} capacity, outperforming their 2D counterparts.  The
\mempool{8}{3D} design achieves the highest performance,
\SI{8.4}{\percent} above the baseline.

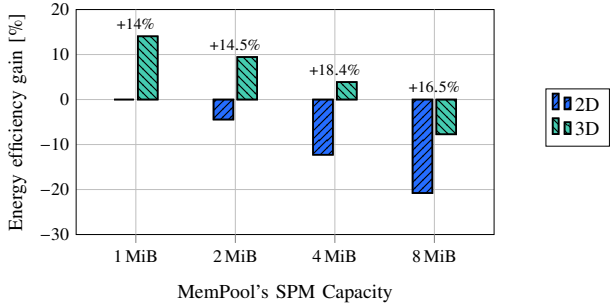
\begin{figure}[htpb]
  \centering
  \resizebox{.9\linewidth}{!}{
  \begin{tikzpicture}
    \begin{axis}[
      ybar,
      height=5.6cm,
      width=\linewidth,
      enlargelimits=0,
      scaled y ticks = false,
      legend style={
        at={(1.2,0.52)},
        anchor=center,
        legend columns=1
      },
      legend cell align={left},
      ylabel={Energy efficiency gain [\si{\percent}]},
      ymajorgrids=true,
      yminorgrids=true,
      minor y tick num=0,
      ytick style={draw=none},
      ymin=-30,
      ymax=20,
      ytick={-30,-20,-10,0,10,20},
      xlabel={MemPool's \gls{SPM} Capacity},
      xmode=log,
      log basis x={2},
      log ticks with fixed point,
      xmajorgrids=true,
      xtick pos=left,
      xtick=\empty,
      extra x ticks={1, 2, 4, 8},
      extra x tick labels={\SI{1}{\mebi\byte}, \SI{2}{\mebi\byte}, \SI{4}{\mebi\byte}, \SI{8}{\mebi\byte}},
      enlarge x limits=0.2]

      \pgfplotstableread{fig/benchmarks/matmul_efficiency.csv}\loadedtable;
      \addplot [fill=color1, thick, postaction={pattern=north east lines}] table [x=Memory, y expr=100*\thisrowno{1}-100] {\loadedtable};
      \addplot [fill=color2, thick, postaction={pattern=north west lines}] table[x=Memory, y expr=100*\thisrowno{2}-100] {\loadedtable};

      \addplot [only marks,
                point meta=explicit,
                nodes near coords={\footnotesize\pgfmathprintnumber[precision=1,showpos]{\pgfplotspointmeta}\%},
                nodes near coords style={above}]
         table [x=Memory,y expr={max(100*\thisrowno{1}-100,100*\thisrowno{2}-100,0)},
                meta expr={(\thisrowno{2}-\thisrowno{1})/\thisrowno{1} * 100}] {\loadedtable};

      \legend{
        2D,
        3D};
    \end{axis}
  \end{tikzpicture}}
\caption{MemPool's energy efficiency gain with larger \gls{SPM}
  capacity, relative to \mempool{1}{2D} with a
  \SI{16}{\byte\per\cycle} off-chip memory bandwidth, when running the
  matrix multiplication kernel. The percentages above the bars
  indicate the energy efficiency gain of the MemPool-3D instance
  compared to the MemPool-2D with the same \gls{SPM} capacity. Higher
  is better.}
  \label{fig:efficiency_analysis}
\end{figure}

The energy efficiency, as expected, shows the opposite trend than the
performance. As shown in \Cref{fig:efficiency_analysis}, the energy
efficiency of the MemPool designs tends to decrease with an increasing
\gls{SPM} capacity. The \mempool{8}{2D} group achieves the worst
energy efficiency, \SI{21}{\percent} below the efficiency of the
\mempool{1}{2D} design. In addition, the 3D designs consistently
outperform their 2D counterparts. For example, the \mempool{4}{3D}
design achieves an efficiency \SI{18.4}{\percent} higher than the 2D
design with the same \gls{SPM} capacity.

\begin{figure}[htpb]
  \centering
  \resizebox{.9\linewidth}{!}{
  \begin{tikzpicture}
    \begin{axis}[
      ybar,
      height=5.6cm,
      width=\linewidth,
      enlargelimits=0,
      scaled y ticks = false,
      legend style={
        at={(1.2,0.52)},
        anchor=center,
        legend columns=1
      },
      legend cell align={left},
      ylabel={\Gls{EDP} variation [\si{\percent}]},
      ymajorgrids=true,
      yminorgrids=true,
      ytick={-20,-10,0,10,20,30},
      minor y tick num=0,
      ytick style={draw=none},
      ymin=-20,
      ymax=30,
      xlabel={MemPool's \gls{SPM} Capacity},
      xmode=log,
      log basis x={2},
      log ticks with fixed point,
      xmajorgrids=true,
      xtick pos=left,
      xtick=\empty,
      extra x ticks={1, 2, 4, 8},
      extra x tick labels={\SI{1}{\mebi\byte}, \SI{2}{\mebi\byte}, \SI{4}{\mebi\byte}, \SI{8}{\mebi\byte}},
      enlarge x limits=0.2]

      \pgfplotstableread{fig/benchmarks/matmul_energy_delay.csv}\loadedtable;
      \addplot [fill=color1, thick, postaction={pattern=north east lines}] table [x=Memory, y expr=100*\thisrowno{1}-100] {\loadedtable};
      \addplot [fill=color2, thick, postaction={pattern=north west lines}] table[x=Memory, y expr=100*\thisrowno{2}-100] {\loadedtable};

      \addplot [only marks,
                point meta=explicit,
                nodes near coords={\footnotesize\pgfmathprintnumber[precision=1,showpos]{\pgfplotspointmeta}\%},
                nodes near coords style={above}]
         table [x=Memory,y expr={max(100*\thisrowno{1}-100,100*\thisrowno{2}-100)},
                meta expr={(\thisrowno{2}-\thisrowno{1})/\thisrowno{1} * 100}] {\loadedtable};

      \legend{
        2D,
        3D};
    \end{axis}
  \end{tikzpicture}}
\caption{MemPool's \gls{EDP} variation with larger \gls{SPM} capacity,
  relative to \mempool{1}{2D} with a
  \SI{16}{\byte\per\cycle} off-chip memory bandwidth, when running the
  matrix multiplication kernel. The percentages above the bars
  indicate the \gls{EDP} variation of the MemPool-3D instance compared
  to the MemPool-2D with the same \gls{SPM} capacity. Lower is
  better.}
  \label{fig:edp_analysis}
\end{figure}
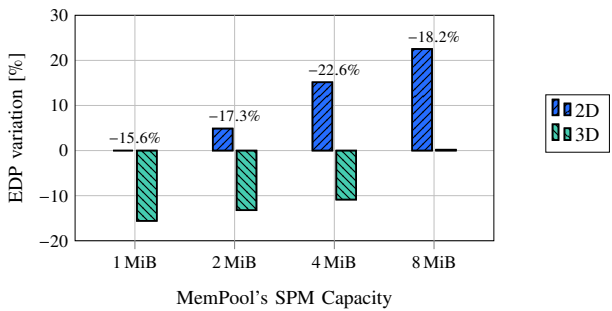

The results of \Cref{fig:efficiency_analysis} indicate that the
\mempool{1}{3D} is the optimal design from the energy efficiency point
of view, \SI{14}{\percent} higher than \mempool{1}{2D}. In addition,
three-dimensional integration allows for the implementation of
\mempool{4}{3D}---a design with four times as much \gls{SPM} capacity
as the baseline design, \mempool{1}{2D}---on an energy budget
\SI{3.7}{\percent} smaller. This intrinsic tradeoff between
performance and energy efficiency can be better analyzed with the
\gls{EDP} results of \Cref{fig:edp_analysis}, the \gls{EDP} defined as
the product of the total energy consumption by the runtime. The
\mempool{1}{3D} configuration has the lowest \gls{EDP},
\SI{15.6}{\percent} below the baseline.

\section{Conclusions}
\label{sec:conclusions}

In this paper, we analyzed the power, performance, area, and energy
efficiency of MemPool, through a co-exploration of its architectural
and technological parameter spaces. We implemented MemPool with
\SIlist{1;2;4;8}{\mebi\byte} of \gls{SPM}, using 2D and 3D
implementation flows, on a modern commercial \SI{28}{\nano\meter}
technology node.

We explored MemPool's performance while running a large matrix
multiplication kernel as a function of the L1 \gls{SPM} capacity and
the off-chip memory bandwidth. For a realistic bandwidth of
\SI{16}{\byte\per\cycle}, we observe a cycle count reduction of
\SI{16}{\percent} when increasing the \gls{SPM} capacity from
\SIrange{1}{8}{\mebi\byte}.

Despite the maximum operating frequency degradation with an increasing
\gls{SPM} capacity, the 3D designs can still achieve an operating
frequency up to \SI{9.1}{\percent} higher than their 2D counterparts.
The \mempool{8}{3D} design achieves a performance \SI{8.4}{\percent}
higher than the \mempool{1}{2D} baseline. The 3D designs consistently
outperform their 2D counterparts by up to
\SI{9.1}{\percent}. Regarding energy efficiency, the 3D designs
outperform their 2D counterparts by up to \SI{18.4}{\percent}. While
increasing the \gls{SPM} size in the 2D case leads to worse energy
efficiency, all but the largest 3D designs achieve a better energy
efficiency than the 2D baseline. We are able to implement the
\mempool{4}{3D} design with an energy budget \SI{3.7}{\percent}
smaller than the 2D instance with only one-fourth of the \gls{SPM}
capacity, \mempool{1}{2D}. To summarize, in this paper we showed the
need for a co-exploration approach with full 3D implementations to
optimize modern designs constrained by their interconnect subsystems.

\bibliographystyle{IEEEtran} \bibliography{mempool_3d}

\end{document}